\documentclass[a4paper]{article}
\usepackage{graphicx}
\usepackage{titlesec}
\titleformat*{\section}{\large\bfseries}

\pdfoutput=1    

\def \ShortTitle#1{\def\@ShortTitle{#1}}
\def \FullConference#1{\def\@FullConference{#1}}

\renewcommand{\abstract}[1]
	{\gdef\abstract{#1}}

\usepackage{amsmath}
\usepackage{bm}		
\newcommand{\dif}[1]{\mathrm{d} #1 \,}

\newcommand{\MeVc}{MeV\kern -0.15em $/\kern -0.08em c$}
\newcommand{\MeVcc}{MeV\kern -0.15em $/\kern -0.08em c^2$}
\newcommand{\GeVc}{GeV\kern -0.15em $/\kern -0.08em c$}
\newcommand{\GeVcc}{GeV\kern -0.15em $/\kern -0.08em c^2$}

\newcommand{\phiH}{\phi_\mathrm{h}}

\newcommand{\kT}{\vec{k_\mathrm{T}}}

\newcommand{\pT}{\vec{p}_{\kern -0.15em \perp}}
\newcommand{\pTsq}{p_{\kern -0.15em \perp}^2}

\newcommand{\PhT}{P_{h\mathrm{T}}}
\newcommand{\PhTt}{$P_{h\mathrm{T}}$}

\newcommand{\A}[2]{A_\mathrm{#1}^{#2}}
\newcommand{\refeq}[1]{Eq.~(\ref{#1})}
\newcommand{\mr}[1]{\mathrm{#1}}
\renewcommand{\vec}[1]{\bm{#1}}

\title{Measurement of the azimuthal modulations of hadrons in~unpolarised SIDIS}

\ShortTitle{Measurement of the azimuthal modulations of hadrons in~unpolarised SIDIS}

\author{\speaker{J.~Matou\v{s}ek}, on behalf of the COMPASS Collaboration\\
        University and INFN Trieste, Italy\\
        E-mail: \email{jan.matousek@cern.ch}}

\abstract{
In 2016 and 2017 COMPASS has collected a considerable amount of deep inelastic scattering events with a 160~\GeVc\ muon beam and a liquid hydrogen target. An analysis of 4\,\% of these data has allowed to extract preliminary results on the amplitudes of three azimuthal modulations, $\A{UU}{\cos\phiH}$, $\A{UU}{\cos2\phiH}$ and $\A{LU}{\sin\phiH}$. The first two are particularly important since they carry information on the intrinsic transverse momentum $\kT$ of the quarks and on the correlations between the quark spin and $\kT$, expressed by the Boer--Mulders TMD PDFs. They show strong kinematic dependence as a function of the Bjorken variable $x$, of the fraction of virtual photon energy carried by the hadron $z$ and of the component \PhTt\ of the hadron momentum orthogonal to the virtual photon direction, thus confirming the results of previous measurements. 

Also, as was recently shown by COMPASS, hadrons coming from decay of diffractively produced vector mesons give significant contribution to the amplitudes in certain kinematic regions, explaining in part the kinematic dependencies and helping the interpretation of the measurements. The effect on the published COMPASS $^6$LiD data is discussed.
}

\FullConference{XXVII International Workshop on Deep-Inelastic Scattering 
		and Related Subjects - DIS2019\\
		8-12 April, 2019\\
		Torino, Italy}

\newcommand{\speaker}[1]{#1}
\newcommand{\email}[1]{{\normalsize \texttt{#1}}}
\date{}

\begin{document}

\maketitle

\vspace{-15pt}
\parbox{.9\textwidth}{	
    \begin{center}
        Presented at \textit{\let\\=\relax \@FullConference}
    \end{center}
    }

\centerline{
    \parbox{.9\textwidth}{	
    \begin{center}
        {\bf Abstract}
    \end{center}
    \vspace{-5pt}
    \abstract}
    }

\section{Introduction}

\begin{figure}[t]
	\centering
    \includegraphics[width=0.33\textwidth]{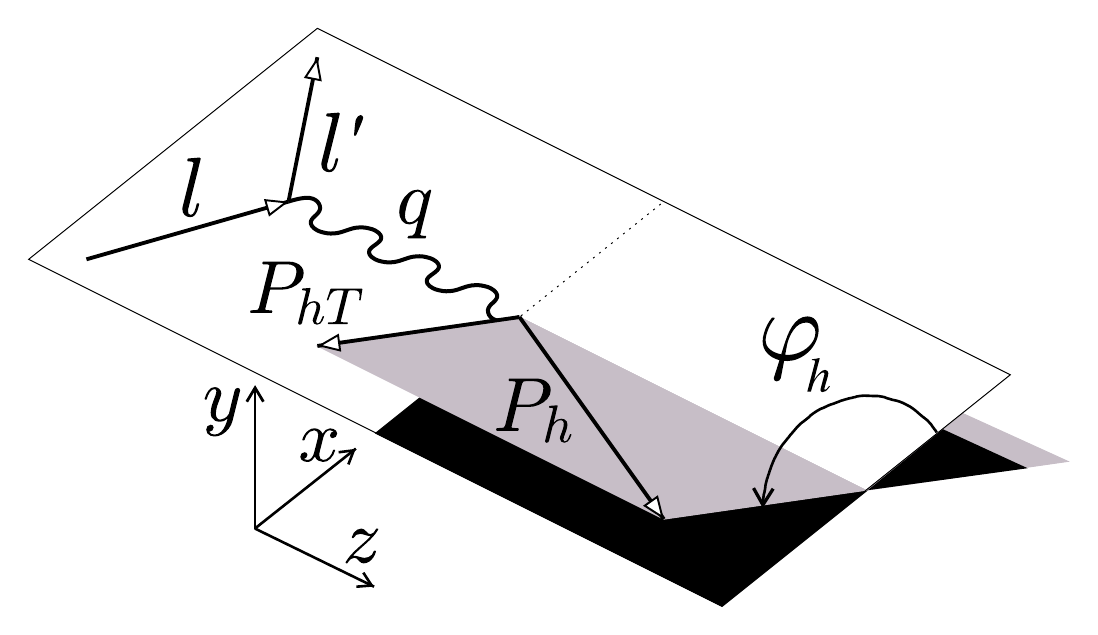}
    \caption{\label{fig:gammaN} SIDIS in the $\gamma^*$--$N$ system. $l$ and $l^\prime$ are the lepton momenta, $\vec{P}_h$ is the momentum of the observed hadron, $\vec{P}_{h\mr{T}}$ is its transverse component with respect to the $\gamma^*$ direction and $\phiH$ its azimuthal angle.}
\end{figure}

The production of hadrons in the semi-inclusive deep inelastic scattering (SIDIS) of leptons off a fixed target,
$
	\ell N \rightarrow \ell^\prime h X
$
is a powerful tool for probing of the nucleon structure. The differential cross section in the one-photon exchange approximation can be written as~\cite{bacchetta:2007}
\begin{align}
	\label{eq:xsec}
	\frac{\dif\sigma}{\dif{x}\dif{y}\dif{z}\dif{\PhT^2}\dif{\phi_h}}
    = \sigma_0\left( 1 + \epsilon_1 \A{UU}{\cos\phi_h} \cos\phi_h
    	+ \epsilon_2 \A{UU}{\cos2\phi_h} \cos2\phi_h
        + \lambda \epsilon_3 \A{LU}{\sin\phi_h} \sin\phi_h \right),
	\\
	\mr{where}
	\quad
	\epsilon_1 = \frac{2(2-y)\sqrt{1-y}}{1+(1-y)^2},
	\qquad
	\epsilon_2 = \frac{2(1-y)}{1+(1-y)^2},
	\qquad
	\epsilon_3 = \frac{2y\sqrt{1-y}}{1+(1-y)^2}.
\end{align}
Here $\PhT$ and $\phiH$ are the transverse momentum and the azimuthal angle of the hadron $h$ (see Fig.~\ref{fig:gammaN}), $z$ is the fraction of the photon energy carried by the hadron, $y$ and Bjorken $x$ are the usual DIS variables, $\sigma_0$ is the $\phiH$-independent cross-section, $\lambda$ is the beam longitudinal polarization\footnote{The beam used by COMPASS is longitudinally polarised positively for $\mu^-$ and negatively for $\mu^+$ by about 80\,\%.} and the amplitudes $A_X^Y$ of the azimuthal modulations carry information on the internal structure of the proton and on quark fragmentation. They have been studied by COMPASS in the production of charged hadrons in the scattering of muons off a $^6$LiD (effectively isoscalar) target~\cite{COMPASS:2014pwc}. The results have been provided as a function of $x$, $z$ and $\PhT$ simultaneously (3D) and integrated over two of the variables (1D). Preliminary 1D results from a hydrogen target have been recently presented~\cite{Moretti:2019lkw} and are discussed in Sec.~\ref{sec:2016}. The asymmetries measured on both isoscalar and hydrogen targets show strong kinematic dependences, something that has been observed by HERMES as well~\cite{Airapetian:2012yg}.

The $\cos\phiH$ and the $\cos2\phiH$ azimuthal asymmetries receive contributions at twist three and four respectively by the Cahn effect, which originates from the kinematics, when the intrinsic quark momentum $\kT$ is considered. Thus it gives access to $\langle k_\mr{T}^2 \rangle$, in a complementary way to the \PhTt-dependent hadron multiplicities in SIDIS, which are studied by COMPASS as well\,\cite{COMPASS:2017ctw,Moretti:2019dis}. The Boer--Mulders transverse momentum dependent parton distribution functions (TMD PDFs), which describe the transverse polarisation of quarks in an unpolarised hadron, are a leading twist contribution to the $\cos2\phi$ modulation, and a twist three contribution to the $\cos\phiH$ modulation, in both cases coupled with the Collins fragmentation functions. The $\sin\phiH$ asymmetry is attributed to higher-twist effects and has no intuitive interpretation in the parton model. Several phenomenological analyses of the results, e.g. Ref.~\cite{Barone:2015ksa}, did not succeed to reproduce the data. Thus the knowledge of the $\langle k_\mr{T}^2 \rangle$ so far relies on the multiplicities, which probe only the combination of the $\kT$ and of the transverse momentum obtained in the fragmentation. Possible non-zero Boer--Mulders PDF still remains to be demonstrated in SIDIS.

It is known (e.g. Ref.~\cite{COMPASS:2017ctw}) that part of the hadron sample used for COMPASS multiplicity and azimuthal asymmetry analyses originates from the diffractive production of vector mesons ($\rho^0$, $\phi$, $\omega$) that decay into lighter hadrons. This process, which can be described by the fluctuation of the virtual photon into a vector meson that interacts with the nucleon via multiple-gluon exchange, can not be taken into account in the usual TMD PDF interpretation outlined above. Recently~\cite{Kerbizi:2018bvk} it has been shown that the $\pi$ and K from these decays are produced with large and strongly kinematic-dependent $\cos\phiH$ and $\cos2\phiH$ azimuthal modulations. As a result, they contribute significantly to the $\A{UU}{\cos\phi_h}$ and $\A{UU}{\cos2\phi_h}$. These contributions have been estimated and subtracted for the case of the $^6$LiD target, as discussed in the next section. More details can be found in Ref.~\cite{Kerbizi:2018bvk}.
\section{The contribution from diffractive vector meson decays 
		to the $^6$LiD measurement}
\label{sec:dvm}
The fraction of the final-state hadrons originating from the diffractive vector meson (DVM) decays in the SIDIS hadron sample has been estimated in an earlier work~\cite{COMPASS:2017ctw} using two Monte Carlo (MC) simulations: LEPTO to generate SIDIS events and HEPGEN~\cite{Sandacz:2012at} for the diffractive $\rho^0$ and $\phi$ events. To get the fraction in the bins of the azimuthal asymmetries, which are different from Ref.~\cite{COMPASS:2017ctw}, a parametrisation has been used. Only the $\rho^0$, giving the dominant contribution, has been considered. The resulting fraction $r(x,z,\PhT)$ reaches its maximum of about 0.5 in the smallest $x$, largest $z$ and smallest $\PhT$ bin and then decreases monotonously (see Fig.~6 in Ref.~\cite{Kerbizi:2018bvk}). The systematic uncertainty on $r$ is 30\,\% due to the theoretical diffractive cross-section.

The azimuthal modulations of the hadrons coming from the DVM decays have been measured. First a hadron sample has been selected with the same cuts as for the SIDIS asymmetries. Then only the events with two oppositely charged hadrons with the total energy fraction $z_\mr{t} = z_1 + z_2 > 0.95$ were kept in the sample. It can be assumed that the hadrons from DVMs dominate such a sample of exclusive events. Finally, the distributions of $\phiH$ in each one of the 3D bins have been corrected for the acceptance and fitted with
\begin{equation}
	N_\mr{excl}(\phiH,x,z,\PhT) = a_0 \bigl[ 1 
		+ \epsilon_1 a_\mr{UU}^{\cos\phiH}(x,z,\PhT) \cos\phiH
		+ \epsilon_2 a_\mr{UU}^{\cos2\phiH}(x,z,\PhT) \cos2\phiH \bigr].
\end{equation}
The resulting amplitude $a_\mr{UU}^{\cos\phiH}$ decreases with $\PhT$. It is positive for $z < 0.4$, reaching more than 0.5 for small $\PhT$, then it changes sign and becomes negative for $z > 0.55$, reaching almost $-0.5$. The $a_\mr{UU}^{\cos2\phiH}$ amplitude is smaller but still non-negligible with respect to the measured SIDIS asymmetry. Both modulations are almost equal for positive and negative hadrons, indicating that the effect comes from the modulation of the parent vector meson direction.
\begin{figure}[p]
	\centering
    \includegraphics[width=0.7\textwidth]{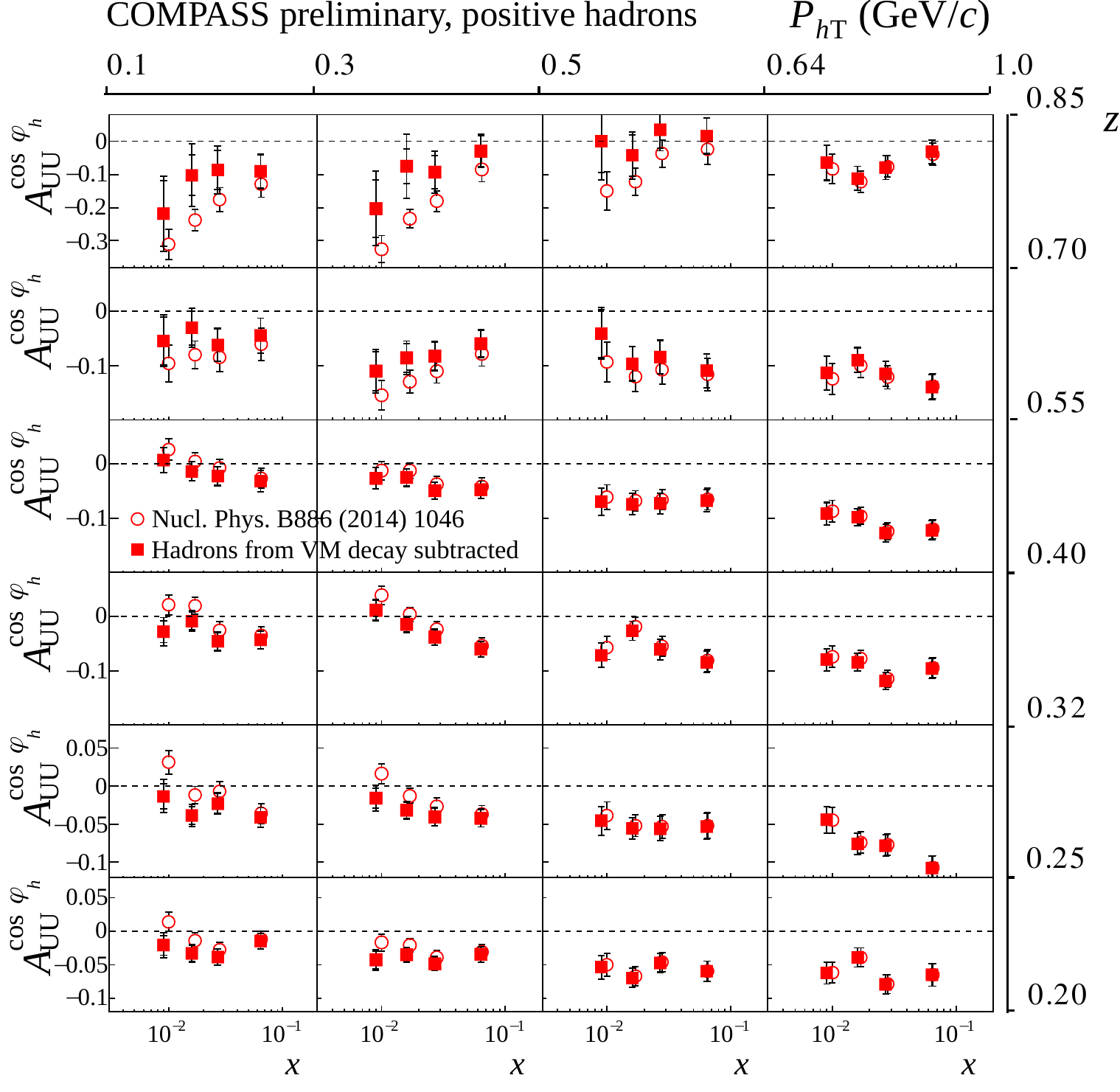}\\
	\vspace{5pt}
    \includegraphics[width=0.7\textwidth]{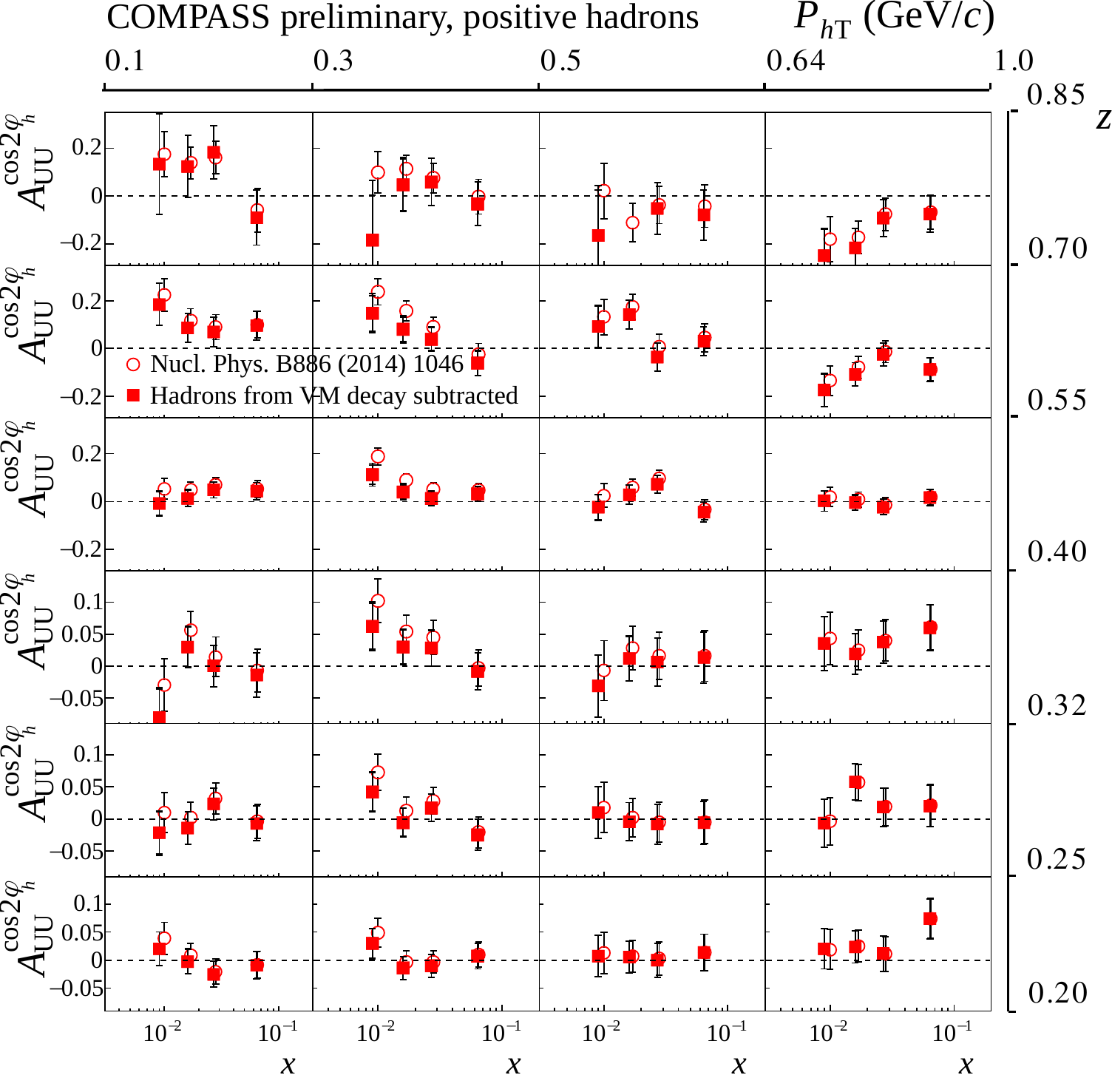}
    \caption{\label{fig:asymsub} 
		The $\cos\phiH$ and the $\cos2\phiH$ asymmetries in SIDIS on 
		$^6$LiD target~\cite{COMPASS:2014pwc}
		for positive hadrons before (empty points) and after (full points) subtraction
		of the contribution coming from the decay of diffractive vector mesons
		done in Ref.~\cite{Kerbizi:2018bvk}. Only statistical uncertainties are shown. 
		The impact on negative hadrons is similar and is not shown due to 
		space limitations.}
\end{figure}
The contributions to the measured azimuthal asymmetries in SIDIS from the hadrons from DVMs are given by the contamination $r$ and the amplitudes $a_\mr{UU}^{\cos\Phi}$ and can be subtracted as
\begin{equation}
	\A{UU}{\cos\Phi}\bigl|_\mr{DVM\,sub.}
		= \frac{\A{UU}{\cos\Phi} - r \, a_\mr{UU}^{\cos\Phi}}{1-r},
	\qquad
	\Phi = \phiH, 2\phiH.
\end{equation}
The asymmetries before and after the subtraction are shown in Fig.~\ref{fig:asymsub}. The effect on the $\A{UU}{\cos\phiH}$ is especially significant: it becomes smoother and almost all positive values, which are difficult to be explained by the Cahn effect, are shifted to zero or below zero. Recently, a newly developed MC describing the fragmentation of polarised quarks~\cite{Kerbizi:2018qpp}, modified to include the Cahn effect, has been used to describe 	the $\A{UU}{\cos\phiH}$ asymmetry. Subtracting the contribution from the DVM decay hadrons has considerably improved the description of the data~\cite{Kerbizi:2018bvk}. The effect will be studied in detail in the case of the new hydrogen target data, even if the DVM contribution is expected to be somewhat smaller on proton.
\section{Measurement on liquid hydrogen target}
\label{sec:2016}
A large sample of SIDIS events on protons has been collected by COMPASS in 2016 and 2017 with alternating $\mu^+$ and $\mu^-$ beams at 160\,\GeVc\ and a liquid hydrogen target. The preliminary results presented in Ref.~\cite{Moretti:2019lkw} and here are based on about 4\,\% of the full data set. The event selection has been identical to the $^6$LiD target measurement. In particular, we have asked for the squared photon virtuality $Q^2>1~(\mr{GeV}/c)^2$, mass of the final hadronic state $W>5$~\GeVcc, $0.2<y<0.9$ and $0.003<x<0.13$. To stay in the current fragmentation region and to reduce exclusive events only hadrons with $0.2<z<0.85$ have been used, while $0.1~(\mr{GeV}/c) < \PhT < 1~(\mr{GeV}/c)$ ensures good $\phiH$ resolution and small acceptance correction. The selected sample consists of almost 270\,000 positive and 215\,000 negative hadrons with the $\mu^+$ beam; for the $\mu^-$ beam it is about 10\,\% less.

\begin{figure}[t]
	\centering
    \includegraphics[width=0.68\textwidth]{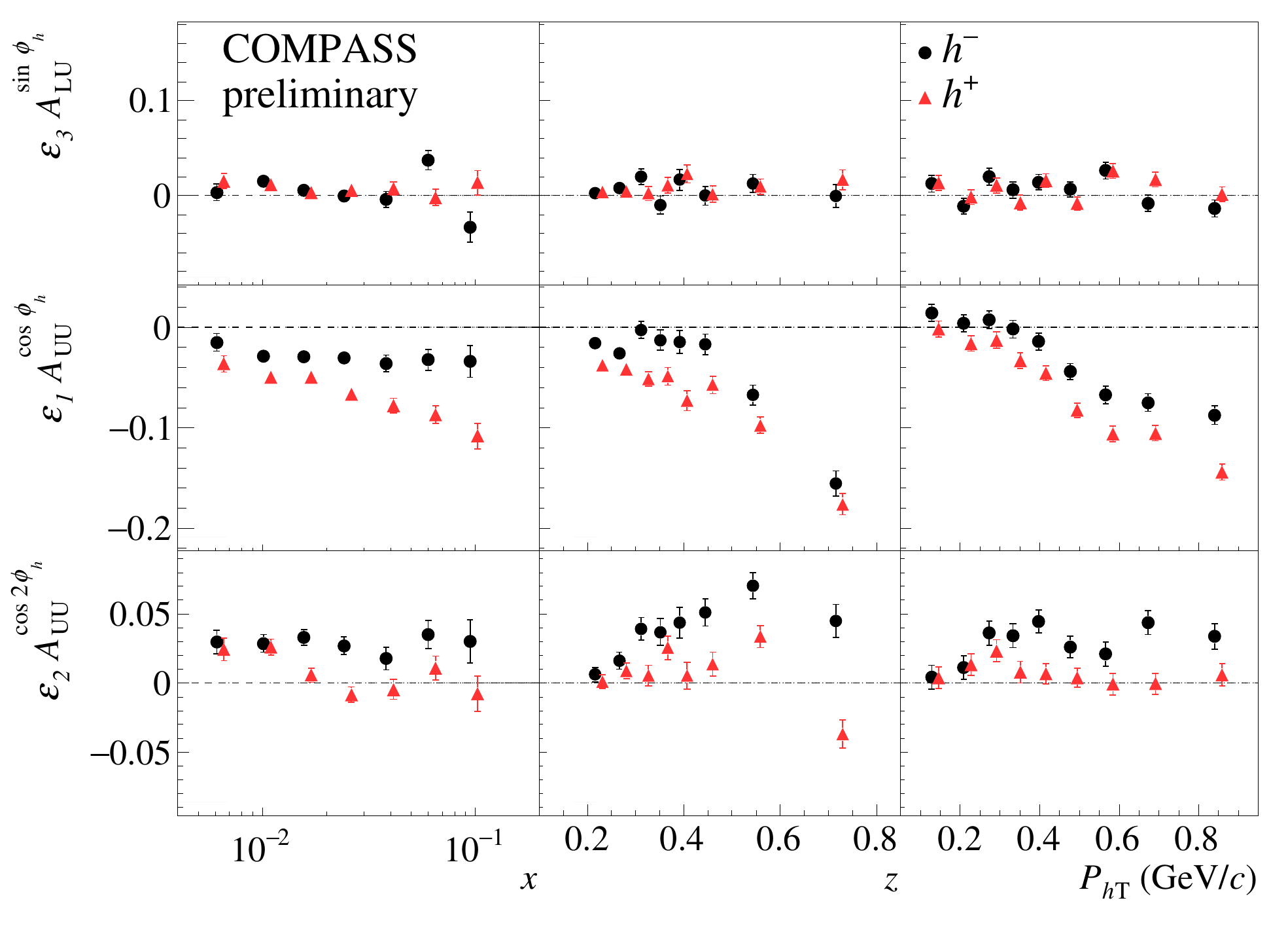}
	\vspace{-10pt}
    \caption{\label{fig:asym2016} Azimuthal asymmetries for charged
		hadrons produced in SIDIS off a liquid hydrogen target~\cite{Moretti:2019lkw}.
		Only statistical uncertainties are shown.}
\end{figure}
The hadrons have been divided into kinematic bins of $x$, $z$ and $\PhT$ separately (1D). The distributions of $\phiH$ in each bin have been corrected for the acceptance, obtained from a MC based on LEPTO generator and simulating the full setup and reconstruction. The acceptance corrections are smaller than 10\,\% in every bin. Finally, the distributions have been fitted with \refeq{eq:xsec}. The central region $-\pi/8 < \phiH < \pi/8$ has been removed from the fit, because it is contaminated by electrons coming from radiative photons that are emitted along the direction of the muons. We have done the procedure separately for the $\mu^+$ and $\mu^-$ beam and then, the results being compatible as expected, we have combined them. The results are shown in Fig.~\ref{fig:asym2016}. While $\epsilon_3 \A{UU}{\sin\phiH}$ has a flat trend, the other two asymmetries show strong kinematic dependencies, in particular $\epsilon_1 \A{UU}{\cos\phiH}$ reaches absolute value of almost 20\,\% at high $z$. These trends closely resemble those observed on the $^6$LiD target~\cite{COMPASS:2014pwc}.


\section{Conclusions}
A large sample of SIDIS events has been collected by COMPASS in 2016 and 2017 with  a 160\,\GeVc\ muon beam and hydrogen target. 4\,\% of the full data set have been used to provide preliminary azimuthal modulations of charged hadrons as a function of $x$, $z$ and $\PhT$ separately. The observed trends are similar to the SIDIS asymmetries at the $^6$LiD target. The current analysis will be extended to the full sample and studies are ongoing to access a wider kinematic range. Multi-dimensional binning including $Q^2$ is foreseen as well.

COMPASS has measured the azimuthal modulations of charged hadrons coming from the decay of vector mesons produced diffractively in the scattering of 160~\GeVc\ muons off a $^6$LiD target. The contribution from these hadrons to the published COMPASS $\A{UU}{\cos\phiH}$ and $\A{UU}{\cos2\phiH}$ azimuthal asymmetries has been quantitatively estimated and shown non-negligible over almost all the explored kinematic range. The work to estimate them for the new hydrogen target data is ongoing. Subtracting this contribution is expected to help the phenomenological analyses, which try to interpret the asymmetries in the TMD framework extracting the Boer--Mulders TMD PDF and the intrinsic quark momentum.

\providecommand{\href}[2]{#2}\begingroup\raggedright\endgroup

\end{document}